# COMMUNICATION

## Electromechanical dopant-defect interaction in acceptor-doped ceria

Ahsanul Kabir[a], Victor Buratto Tinti[a], Maxim Varenik[b], Igor Lubomirsky[b], Vincenzo Esposito[a]*

**Oxygen defective cerium oxides $CeO_{2-\delta}$ exhibits a non-classical giant electromechanical response that is superior to lead-based electrostrictors. In this work, we report the key-role of acceptor dopants, with different size and valence ($Mg^{2+}$, $Sc^{3+}$, $Gd^{3+}$, and $La^{3+}$), on polycrystalline bulk ceria. Different dopants tune the electrostrictive properties by changing the electrosteric dopant-defect interactions. We find two distinct electromechanical behaviors: when the interaction is weak (dopant-vacancy binding energy ≤ 0.3 eV), electrostriction displays high coefficient ($M_{33}$), up to $10^{-17}$ $(m/V)^2$, with strongly time-dependent effects. In contrast, we observe no time-dependent effects when the interaction becomes strong (≥ 0.6 eV).**

Over the last four decades, oxygen defective ceria has been widely used in multiscale applications from electrochemical devices to automotive catalysts, gas sensors, *etc.*[1,2] Furthermore, a recent discovery demonstrates that ceria thin films display an unconventional electromechanical strain at room temperatures.[3,4,5,6] For example, the reported electrostriction coefficient ($M_e$) is around $6.5 \cdot 10^{-18}$ $(m/V)^2$ for $[V_O^{..}]$ = 5%, *i.e.* 20 mol% Gd-doped ceria (GDC).[3] Such a value is extremely high for a material with low dielectric constant ($\varepsilon_r^{GDC} \approx 30$)[7] and high elastic modulus ($Y^{GDC} \approx 200$ MPa).[8] Surprisingly, in contrast to conventional inorganic ceramics material, ceria thin films expand perpendicularly to the applied field direction. These properties are also investigated in bulk ceria and other oxygen defective fluorites ($Bi_2O_{3-\delta}$), illustrating similar outcomes.[9,10,11,12,13,14] The atomistic origin of this uncommon behavior is associated with the presence of charge compensating oxygen vacancies ($V_O^{..}$) in the ceria lattice.[15,16] The $V_O^{..}$ creates an electroactive elastic dipole ($Ce_{Ce}$-

$V_O^{..}$) having a long bond length and anomalous six reduced ($Ce_{Ce}$-$O_O$) bonds when compared to the bond length of undoped ceria, leading to the development asymmetric charge distribution and anisotropic local dipolar elastic field around the host lattice.[17] Upon interaction with an electric field, such distorted fluorite lattice ($Ce_{Ce}$-$O_O$-$V_O^{..}$) complex turns to be more ideal-like fluorite structure and successively induces a considerable local strain and vice-versa.[15,16] Although governed by the $Ce_{Ce}$-$V_O^{..}$ complex, the local configuration of oxygen vacancy is a crucial feature controlling electrostriction in bulk ceria.[15,18] Literature reports of both computational and experimental works demonstrate that the configuration of oxygen vacancy in the lattice strongly depends on dopant size and valence, *i.e.*, electrostatic (attractive) and elastic (repulsive) interaction between the cation and vacancy.[19,20] **Fig. 1a** schematically represents a simple configuration of dopant cation concerning nearby oxygen vacancies in a unit cell of ceria. In general, the acceptor dopant prefers to occupy the next neighbour (1nn) or next-nearest neighbour (2nn) lattice site of oxygen vacancy. The electrostatic interaction favors 1nn site while elastic relaxation favors 2nn positions. When the total (electrosteric) dopant-defect interaction is significantly high, dopant tends to cluster with $V_O^{..}$, successively forming small/large scale defect associations and/or segregates locally at the crystalline disorders. The defect associate acts as an electric dipole, in which the binding energy of defect association dominates the ionic conductivity and dielectric constant since it traps more oxygen vacancies.[21]

In this work, we intended to experimentally investigate the effect of dopant type, *i.e.*, total dopant-defect interaction, on the electro-chemo-mechanical properties in bulk ceria. Four different dopants ($M^n$), including $Sc^{3+}$, $Mg^{2+}$, $Gd^{3+}$, and $La^{3+}$, were selected for this

a. Department of Energy Conversion and Storage, Technical University of Denmark, Kgs. Lyngby 2800, Denmark

b. Department of Materials and Interfaces, Weizmann Institute of Science, Rehovot 761001, Israel

†Corresponding Authors: A. Kabir (ahsk@dtu.dk), V. Esposito (vies@dtu.dk)

purpose. The computational calculation in **Fig. 1b** estimates that these dopants have a large span in the binding energy (1nn) values covering both strong and weak dopant-vacancy interactions, *e.g.,* 0.15-1.25 eV.[22] As can be seen, the binding energy of dopant defect interaction highly depends on the dopant type. For example, the Sc- and Mg-doped system introduces a stronger interaction than of Gd- and La-doped ceria. Finally, we correlate the electrostriction behavior with the electrochemical properties from low to intermediate temperatures.

Nanoscale ceria powders with a composition of $Ce_{1-x}M_xO_{2-\delta}$ where x = 0.10 and 0.05 for trivalent (M = $Sc^{3+}$, $Gd^{3+}$,$La^{3+}$) and divalent (M = $Mg^{2+}$) dopant, respectively were prepared using a combination of co-precipitation and nitrate route.[23] At first, the appropriate amount of metal nitrate salts were dissolved in boiling nitric acid (5 N) under mild stirring and slowly cooled. Then pure ceria ($CeO_2$) powder, as synthesized by the co-precipitation method described in Ref. 16 was dispersed in ethanol. Sequentially, a metal nitrate solution was transferred into the suspension. After that, the chemically modified suspension was ball-milled with zirconia balls (2 mm diameter) for 10 hours at 50 rpm, oven-dried at 100 °C, and finally calcined at 500 °C for 2 hours. Next, the powder was softly crushed with agate mortar and pestle and sieved using a 150 μm mesh. For the pellet preparation, the powders were uniaxially cold-pressed (12 mm diameter) under a pressure of 200 MPa for half a minute and sintered at 1450 °C for 10 hours in atmospheric air. The apparent density of the resultant pellet was measured by the Archimedes method in the water medium. The crystallographic phase and microstructure were characterized by X-ray powder diffraction (XRD, Bruker D8) and scanning electron microscopy (SEM, Zeiss Merlin), respectively. For the dielectric and electrochemical measurements, silver paste (electrode) was brushed onto the parallel face of the sample and dried at 600 °C. The electrochemical experiments were performed using two probe method with a frequency response analyzer (FRA-EIS, Solartron 1260) at 250-450 °C in a frequency range from 1 Hz to 10 MHz in the air with an alternating potential of 100 mV. The room temperature impedance (dielectric) measurement was carried out with a dielectric analyzer (Novocontrol) in high voltage mode (10 VAC) with frequencies between 1 MHz-1mHz. The electromechanical strain was estimated at room temperature under ambient condition (relative humidity 25-50%) with a proximity sensor of the capacitive type with lock-in detection, as explained in detail elsewhere.[9,24] The samples were polished (up to 4000 mesh), washed with ethanol/acetone, and then thermally treated at 500 °C for 5 hours in the air. Two mechanically contact metal electrodes made of stainless steel metal plates were used.

As measured, the sintered pellets are highly dense and above 95% of the theoretical density, agreeing with the microstructural analysis (**Fig. S1**). The grains are thermodynamically relaxed and have an average size between 3-5 μm, regardless of the composition. The XRD pattern of the samples corresponds with the reference profile of pure ceria (**Fig. S2**), illustrating a single-phase cubic fluorite structure without any additional phases.

The dielectric constant of a material has a significant role in controlling electrical properties, *i.e.,* dopant-defect interaction and defect cluster.[25] The distinction of dielectric constant ($\varepsilon'$) concerning applied frequency is estimated from the imaginary part ($Z''$) of the complex EIS spectra and represented in **Fig. 2a** at room temperatures. As can be seen, $\varepsilon'$ is relatively constant from high to intermediate frequencies (up to ~10 kHz) for all samples. However, $\varepsilon'$ increases significantly at lower frequencies, which is associated with increased capacitive effect, leading to accumulation of charge at the high-energy barrier site.[26] As estimated, the GDC and LDC sample have lower $\varepsilon'$ value than of SDC and MDC compounds (see **Table 1**), indicating that electrostatic interaction between dopant-defect is small in the former. Characteristically, the dielectric constant increases with increasing temperature (**Fig. S3**), since the polarization of charge carriers raises at higher temperatures. From the EIS spectra, the electrical conductivity of the sample is measured using the Nyquist formalism, which allows separation of grain, grain boundary, and electrode contribution (**Fig. S4 and S5**). The resistivity of all compounds is strongly dominated by the grain boundary and have an ion-blocking factor above 0.9 at 300 °C. The effect of ion blocking property is qualitatively estimated by the parameter called grain boundary blocking factor ($\alpha_{g.b.}$), where $\alpha_{g.b.} = \frac{R_{g.b.}}{R_{bulk}+R_{g.b.}}$.[18] In cerium based oxides, not only grain boundary but also dopant-defect association, nanoclusters, dopant segregation, residual pore all combined introduce ion-blocking effect.[2,27] The temperature-dependent total electrical conductivity ($\sigma$) is illustrated in **Fig. 2b** that follows the Arrhenius expression:

$$\sigma = \sigma_0 \exp\left(-\frac{E_a}{kT}\right) \qquad (1)$$

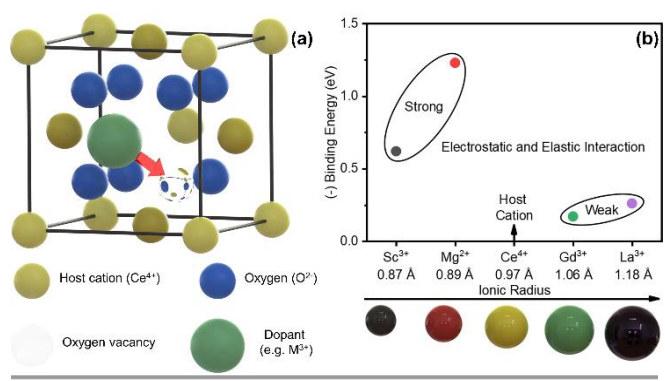

**Figure 1:** (a) Schematic illustration of a unit cell of fluoriute structured ceria, showing electrostreic interaction between acceptor dopant and charge compensating oxygen vacancy. (b) Computationally calculated binding energy (1nn) values of variously doped ceria as a function of dopant radius. The data is taken after V. Butler et al.[22]

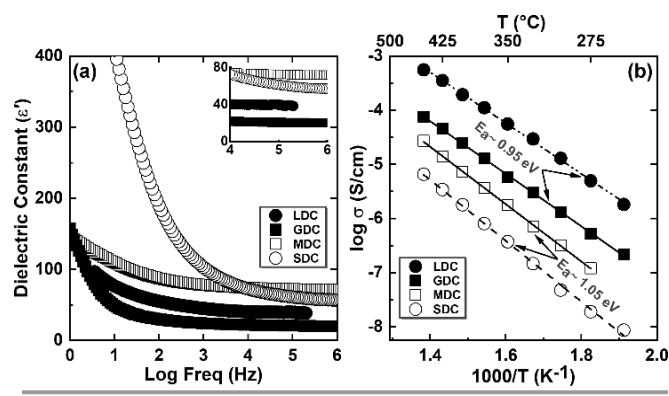

**Figure 2:** (a) Estimation of dielectric constant with respect to frequency, measured via EIS at room temperature under AC bias of 10 V. (b) The temperature-dependent Arrhenius plot for the total electrical conductivity of variously doped ceria samples, examined in air.

Where $\sigma_0$ is the pre-exponent factor and $E_a$ the activation energy. Even though all samples contain equivalent oxygen vacancy concentration, the conductivity value differs significantly between the materials. The magnitude of conductivity increases with temperature showing a maximum for LDC and minimum for the SDC sample. Such outcomes simply emphasize the leading contribution of oxygen vacancy configuration, *i.e.*, dielectric constant, dopant-defect interaction, and defect cluster on tuning charge migration properties. We attribute the lowest conductivity value in SDC to its limited solubility in ceria, which is close to 3%.[28] According to Nowick *et al.*, the Sc³⁺ dopant also acts as a scavenger for oxygen vacancies in cerium oxide.[28] Interestingly, the activation energy value is equivalent (~1.0 eV) in all samples, emphasizing that the mobility of oxygen vacancy is considerably different in each sample due to different dopant associated trap for vacancies.

The electrostrictive strain of the samples, with a response to the applied electric field, is illustrated in **Fig. 3**. As noticed, the strain starts to saturate at a particular electric field value, *e.g.*, ~4-5 kV/cm, for the LDC and GDC samples. The saturation of strain is fitted with the empirical equation:[9]

$$u(E^2) = M_{33} \cdot E_{sat}^2 \cdot \left[1 - \exp\left(-\frac{E^2}{E_{sat}^2}\right)\right] \qquad (2)$$

Here, $M_{33}$ is the electrostriction strain coefficient, and $E_{sat}$ is the saturation electric field. Above the saturation point, a further electrostrictive strain is no longer possible, since the elastic dipole comes to be fully aligned with the high electric field.[14] Besides, at high frequencies, the strain value reduced significantly but followed the typical linear relation. On the other hand, neither strain saturation nor reduction was noticed for the SDC and MDC compounds. The electrostrictive strain coefficient ($M_{33}$) as a function of applied frequency is reported in **Fig. 4a**. As observed, the GDC illustrates maximum $M_{33}$ value having one order higher magnitude in the low frequencies ($\leq 2$ Hz). At the low-frequency regime, for instance, below 10 Hz, the value of $M_{33}$ for the GDC and LDC compound is strongly dependent on applied frequency value. The $M_{33}$ empirically follows a frequency-associated relaxation, given by:[9]

$$M_{33}(f) = \frac{M_{33}^0}{\sqrt{1+(\tau.f)^{2+\alpha}}} + M_{33}^\infty \qquad (3)$$

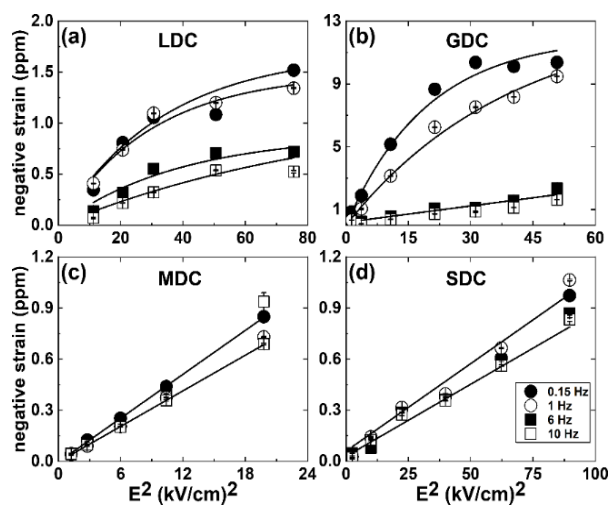

**Figure 3:** (a)-(d) The negative electromechanical strain as a function of electric field square under applied frequencies from ~0.15-10 Hz.

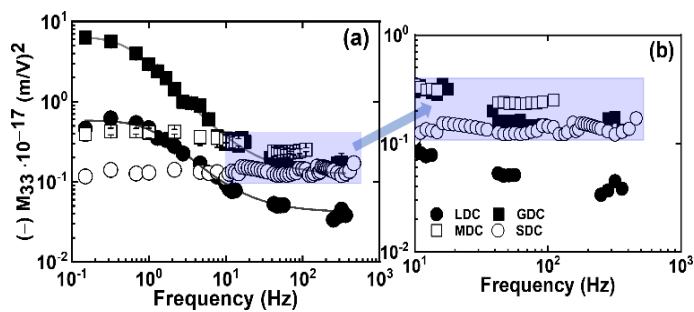

**Figure 4:** Frequency-dependent electrostrictive strain coefficient ($M_{33}$) of all samples at (a) at 0.15-700 Hz (b) high-frequency regime.

where $M_{33}^0$ and $M_{33}^\infty$ are the electrostriction coefficient value at low and high frequencies, respectively. $\tau$ is the characteristic relaxation time, and $\alpha$ is the non-ideality factor. Surprisingly, $M_{33}$ does not obey such relation for the SDC and MDC samples. These materials display a frequency-independent steady $M_{33}$ value about ~0.1-0.3 $\cdot$ $10^{-17}$ (m/V)$^2$. At high-frequency region, $e.g.$, above 10 Hz, the GDC exhibits similar $M_{33}$ value to the former, while LDC displays an order smaller $M_{33}$ of roughly 0.03 $\cdot$ $10^{-17}$ (m/V)$^2$ (see **Fig. 4.b**). Finally, the $M_{33}$ value in the order of $\geq 0.1 \cdot 10^{-17}$ (m/V)$^2$ is still one order of magnitude larger than classical Newnham empirical law suggesting that SDC and MDC compounds can be used for prospective frequency insensitive electrostriction application.

**Table 1:** The estimated dielectric constant, relaxation frequency, and electrostriction coefficient of the samples at room temperatures.

| Sample ID | $f_{g.b.}$ (Hz) | $\varepsilon'$ | $M_{33}$ (1 Hz) (m/V)$^2$ | $M_{33}$ (100 Hz) (m/V)$^2$ |
|---|---|---|---|---|
| **LDC** | 4.0 x $10^1$ | 40 | 0.5 $\cdot$ $10^{-17}$ | 0.04 $\cdot$ $10^{-17}$ |
| **GDC** | 2.0 $\cdot$ $10^1$ | 25 | 3.0 $\cdot$ $10^{-17}$ | 0.15 $\cdot$ $10^{-17}$ |
| **MDC** | 1.0 $\cdot$ $10^1$ | 75 | 0.4 $\cdot$ $10^{-17}$ | 0.25 $\cdot$ $10^{-17}$ |
| **SDC** | 1.0 $\cdot$ $10^2$ | 65 | 0.12 $\cdot$ $10^{-17}$ | 0.15 $\cdot$ $10^{-17}$ |

In this work, variously doped ceria compounds having equivalent oxygen vacancy concentration were fabricated by conventional sintering at higher temperatures. The resultant dense pellets illustrate relaxed grains of micron size. As expected, the samples show a highly deviated dielectric relaxation and total electrical conductivity based on the local configuration of oxygen vacancies built in the materials. Moreover, all these compounds demonstrate an unconventional electrostriction response. Interestingly, the sample having weak dopant-defect interaction in the lattice generates a frequency-dependent electrostriction behavior. Whereas, the material with strong dopant-defect interaction exhibits a frequency-independent steady electrostriction coefficient from low to high frequencies. To sum up, the local configuration of oxygen vacancy configuration and its association with dopant cation is a crucial factor controlling the electromechanical response in cerium-based compounds.

## Conflicts of interest

There are no conflicts to declare.

## Acknowledgment

This research was supported by DFF-Research project grants from the Danish Council for Independent Research, Technology and Production Sciences, grant number 48293 (GIANT-E), and the European H2020-FETOPEN-2016-2017 project BioWings, grant number 801267. The authors would like to thank Rana Al Tahan for her assistance in impedance measurement.